\documentclass{aa}
\usepackage{times}
\usepackage{graphicx}
\usepackage{natbib}

\bibpunct{(}{)}{;}{a}{}{,}

\begin{document}

\title{RXTE Monitoring of Centaurus~A}

\author{S. Benlloch\inst{1}
  \and R.E. Rothschild\inst{2}
  \and J. Wilms\inst{1}
  \and C.S. Reynolds\inst{3,4}
  \and W.A. Heindl\inst{2}
  \and R. Staubert\inst{1}}

\authorrunning{S. Benlloch et al.}  

\offprints{S. Benlloch, \email{benlloch@astro.uni-tuebingen.de}} 

\institute{Institut f\"ur Astronomie und Astrophysik -- Astronomie,
  University of T\"ubingen, Waldh\"auser Str. 64, 72076 T\"ubingen, Germany
  \and Center for Astrophysics and Space Sciences, University of California
  at San Diego, La Jolla, CA 92093, USA \and JILA, University of Colorado,
  Boulder, CO 80309-0440, USA \and Hubble Fellow}

\date{Received $<$date$>$ / Accepted $<$date$>$ }

\abstract{We report on the analysis from $\sim$110\,ks of X-ray
  observations of Centaurus~A carried out with the Proportional Counter
  Array (PCA) and the High Energy X-ray Timing Experiment (HEXTE)
  instruments on Rossi X-ray Timing Explorer (RXTE) during three monitoring
  campaigns over the last 4 years (10\,ks in 1996, 74\,ks in 1998, and
  25\,ks in 2000).  The joint PCA/HEXTE X-ray spectrum can be well
  described by a heavily absorbed power law with photon index 1.8 and a
  narrow iron line due to fluorescence of cold matter.  The measured column
  depth decreased by about $30\%$ between 1996 and 2000, while the detected
  2--10\,keV continuum flux remained constant between 1996 and 1998, but
  increased by $60 \%$ in 2000. Since in all three observations the iron line
  flux did not vary, a corresponding decrease in equivalent width was
  noted.  No appreciable evidence for a reflection continuum in the
  spectrum was detected. We present the interpretation of the iron line
  strength through Monte Carlo computations of various geometries.  No
  significant temporal variability was found in Cen~A at time scales from
  days to tens of
  minutes. %
  }

\maketitle

\section{Introduction}\label{sect:intro}

At a distance of 3.5\,Mpc \citep{hui:93a}, \object{Centaurus~A}
(NGC~5128) is by far the nearest active galactic nucleus (AGN).
Cen~A is a giant double-lobed radio source usually classified as a low
luminosity \citet{fanaroff:74a} Class~I radio galaxy, a misdirected BL~Lac
object \citep{morganti:92a}, a Seyfert~2 object \citep{dermer:95a}, and a
radio-loud AGN viewed from the side ($\sim$$70^{\circ}$) of the jet axis
\citep{ebneter:83a}. Even though the nucleus is heavily obscured (in part
by the famous dust-lane that is seen to cross the galaxy), Cen~A is one of
the brightest extragalactic X-ray sources in the sky and the only AGN where
a high quality spectrum from 0.5\,keV to 1\,GeV can be obtained
\citep{steinle:98a}.  This makes it a valuable laboratory to test our
understanding of the high-energy processes occurring within AGN. A recent
comprehensive review of Cen~A is given by \cite{israel:98a}.

Previous observations have shown Cen~A to have highly complex
X-ray/gamma-ray properties
\citep{turner:97a,doebereiner:96a,feigelson:81a,schreier:79a} with
multi-temperature diffuse flux, a spatially resolved jet in radio and
X-rays \citep{kraft:00a,morganti:99a}, and a nuclear component with complex
low energy absorption and a photon index of $\sim$$1.7$. Iron K-shell
absorption and fluorescent line emission at $\sim$$6.4$\,keV with an
equivalent width of $\sim$$100$\,eV have been observed \citep{turner:97a}.
One unusual spectral feature of Cen~A is that there is no evidence for a
strong Compton reflection component \citep[][found $\Omega / 2\pi <
0.15$]{wozniak:98a}.  This implies there is little cold, Thomson-thick
material in the close vicinity of the AGN that reflects X-rays into our
line of sight. Taken together with the iron line equivalent width, this
result is consistent with transmission of the primary flux through the
line-of-sight absorber and viewing the accretion disk nearly edge-on. On
the other hand, the lack of a reflection component is at odds with the
``Unified Model of AGN'' \citep[e.g.,][Ch.~12]{krolik:99a}, where
Seyfert galaxies do show significant reflection components
\citep{smith:96a}.

\begin{figure*}
  \resizebox{\hsize}{!}{\includegraphics{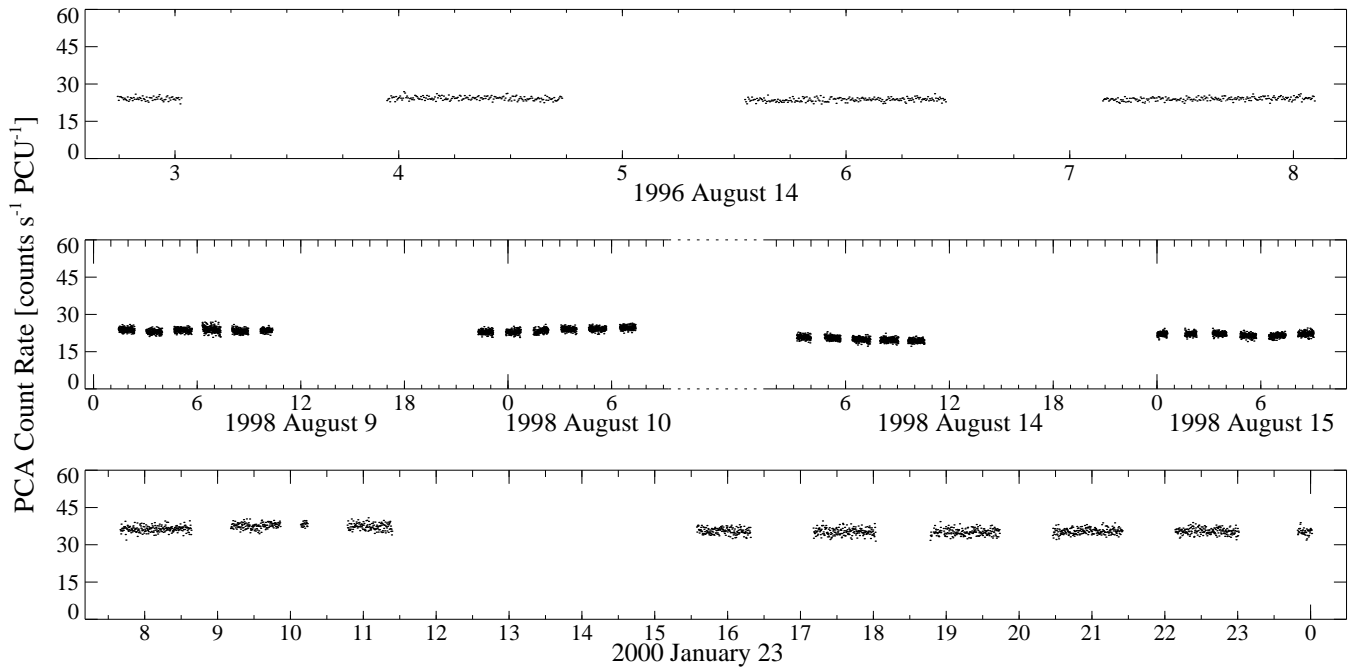}}
  \caption{Background-subtracted PCA-RXTE light curves of Cen~A for the
    three RXTE monitoring campaigns in the years 1996 (upper panel), 1998
    (middle panel) and 2000 (lower panel). The data have a
    resolution of 16\,s and are plotted versus Universal Time (UT) in hours on
    the given date.}\label{fig:pca_lc}
\end{figure*}

In this work we present results of three observations performed by the
Rossi X-ray Timing Explorer (RXTE) -- one of 10\,ks in 1996 August, one of
74\,ks in 1998 August and one of 25\,ks in 2000 January. Preliminary
results from the 1996 August observation, using earlier versions of the
response matrix and background models, have been presented by
\citet{rothschild:99a}. Here we re-analyze these data using the new and
improved PCA response matrices and background models, and present results
on all three observations. The remainder of this paper is structured as
follows. In Sect.~\ref{sec:datareduction} we give the details of the
observations and our data analysis procedure.
Sect.~\ref{sec:dataanalysis} presents the results from the spectral
analysis and the timing analysis. In Sect.~\ref{sec:conclusions} we
discuss our results and summarize the paper.

\begin{table}
  \caption{Details of the RXTE monitoring campaigns of Cen~A}\label{tab:obs}
  \begin{tabular}{cccc}\hline
    Numb. & Obs. Date  & Exposure & Count Rate \\
    & {\footnotesize{(ymd)}} & {\footnotesize{(sec.)}} &
    {\footnotesize{(counts\,$\rm s^{-1}$\, $\rm PCU^{-1}$)}} \\\hline \hline
    1. & 1996, Aug 14     & 10528  & 24 \\
    2. & 1998, Aug 9--15  & 73984  & 22  \\
    3. & 2000, Jan 23 & 24880      & 36  \\
    & \multicolumn{3}{c}{total $\sim$$109.5$\,ks}  \\ \hline
  \end{tabular}\\
  {\footnotesize
    (The count rate is the mean of the 2--20\,keV PCA
    background subtracted count rate)}
\end{table}

\begin{table*}
  \caption{Best Fit Spectral Parameters for the 1996, 1998 and 2000 data
    set fitted separately (Separated spectra) and for the three data sets
    fitted simultaneously (Joined spectrum). }\label{tab:spec2}
    \begin{center}
    \begin{tabular}{cccccccccccc} \hline \hline
      & $ N_{\rm H}$ &  \multicolumn{3}{c}{Iron Line} &
      \multicolumn{2}{c}{Power law} &\multicolumn{2}{c}{Detector
      normalization} &   
      $F_{(2-10\,\rm keV)}$ & $\chi^{2}_{\rm red}$ & d.o.f.  \\

      &              & $E_{\rm Fe}$ &  $A_{\rm Line}$ & EW &
      $\Gamma$  &  $A_{\rm PL}$ &  $A_{\rm Det_{A}}$ & $
      A_{\rm Det_{B}}$ &   \\
       & {\footnotesize {$\times 10^{22}\rm cm^{-2}$}} & {\footnotesize
      {keV}} &
       {\footnotesize {$\times 10^{-4}$}} &
      {\footnotesize {eV}} & &
      {\footnotesize {$\times 10^{-2}$}} & & &
      {\footnotesize {$\times 10^{-2}$}} \\ \hline \hline

      \multicolumn{4}{l}{Separated spectra} & \\
      1996 & 11.30(14) & 6.56(16) &  4.36(74) & 144(26) & 1.92(15) &
      0.135(2) & 0.90(6) & 0.88(9) & 2.28(4) & 0.75 & 64 \\
      1998 & 9.41(28)& 6.47(8) &  3.97(63) & 142(27) &  1.90(2) &
      0.114(4) & 0.92(3) & 0.90(3) & 2.21(4) & 0.51 & 77 \\
      2000 & 7.87(27) & 6.27(9) & 5.46(98) & 120(26) & 1.82(2) &
      0.150(6) & 0.92(3) & 0.92(3) &  3.60(6)& 0.40 & 60 \\ \hline

      \multicolumn{4}{l}{Joined spectrum} &  \\
      1996 &  10.66(18) & 6.49(7)&  5.34(62) & 177(24) &
      1.88(1) & 0.122(3) & {\textit{0.90}} &  {\textit{0.88}} & 2.28(3) &
      0.78  & 209\\  
      1998 & 9.15(19) & 6.45(8) &  4.28(54) & 154(22) & & 0.108(2) &
      {\textit{0.92}} &  {\textit{0.90}} & 2.21(3)  &   \\
      2000 & 8.70(19) & 6.29(11)  & 3.82(87) & 82(20) &  & 0.173(4) &
      {\textit{0.92}} & {\textit{0.92}} &3.58(3)  &   \\ \hline \hline
      
    \end{tabular} \\[0.3cm]
    
  \end{center}
  
  {\footnotesize $E_{\rm Fe}$: line energy. $A_{\rm Line}$: Line
    normalization ($\rm ph\,cm^{-2}\,s^{-1}$ in the line). The Gaussian
    line width was fixed at $\sigma = 0.2\,\rm keV$. EW: line
    equivalent width. $\Gamma$: Photon index of the power law.
    $A_{\rm PL}$: Power law norm ($\rm ph\,keV^{-1}\,cm^{-2}\,s^{-1}$
    at 1 keV). $F_{(2-10\,\rm keV)}$: Flux at the energy range of 2 to
    10\,keV in units of $\rm ph\,cm^{-2}\,s^{-1}$.  
    $A_{\rm Det_{A}}$: Ratio of HEXTE to PCA Normalization for Cluster
    A. $A_{\rm Det_{B}}$: Ratio of HEXTE to PCA Normalization for
    Cluster B. The uncertainties  are 90\% for one parameter
    ($\Delta\chi^2 = 2.7$), and are shown in units of the last digit
    shown. Parameters in italics were frozen.}\\  
\end{table*}

\section{Observations and Data Reduction}\label{sec:datareduction}

There are two pointed instruments on-board RXTE, the Proportional Counter
Array (PCA) and the High Energy X-ray Timing Experiment (HEXTE).  RXTE
observed Cen~A for a total of $\sim$110\,ks over the last five years.  An
observation log is given in Table~\ref{tab:obs}. The PCA light curves are
shown in Fig.~\ref{fig:pca_lc}.

The PCA consists of a set of five co-aligned xenon/methane (with an upper
propane layer) proportional counter units (PCUs) with a total effective
area of $\sim$$7000\,\rm cm^{2}$ . The instrument is sensitive in the
energy range from $2$\,keV to $\sim$$100$\,keV \citep{jahoda:96a}. As the
response matrix is best calibrated between $\sim$2 and 20\,keV, we will
restrict our analysis to this energy range. To increase the signal to noise
ratio we chose only the data from the top xenon layer. Background
subtraction of the PCA data was performed using the ``faint source model''.
To reduce the uncertainty of the PCA background model, we ignore data
measured in the 30 minutes after South Atlantic Anomaly (SAA) passages.
Furthermore, data were not accumulated at times of high electron
contamination as expressed by a certain ratio of veto rates in the
detectors, the so-called ``electron ratio''. We excluded times during which
the ``electron ratio'' was larger than 0.1 in at least one of the
detectors. We accounted for the remaining uncertainty in detector
calibrations by including a systematic error of $1\%$ in the data. See
\citet{wilms:99a} for an in-depth discussion of the PCA calibration issues.

HEXTE consists of two clusters of four NaI(TI)/CsI(Na) phoswich
scintillation counters that are sensitive in the range 15--250\,keV
\citep{rothschild:98a}. Its total effective area is $\sim$1600\,$\rm
cm^{2}$. Background subtraction is done by sequential rocking of the two
clusters on and off of the source position to provide a direct measurement
of the background during the observation. Thus, no calculated background
model is required. HEXTE data between 20\,keV and 200\,keV were used.  To
ensure good statistical accuracy the spectrum was rebinned for fitting, and
no systematic errors were incorporated in the HEXTE data.

The standard data from both instruments were used for the accumulation of
spectra and light curves with 16\,s time resolution. To extract spectra and
light curves, we used the RXTE analysis software in FTOOLS~5.0.  During part
of the observations some PCUs were turned off, and we took this into
account when we added all good PCU data together and combined all
background spectra.  The total PCA response matrix was the average of the
respective estimates for each individual detector combination, weighted by
the fraction of photons measured during the time that such combination was
active. For HEXTE we used the standard response matrices dated 1997
March~20, treating each cluster individually in the data analysis.

On-board RXTE there is a third scientific instrument, the All-Sky Monitor
\citep{levine:96a}. The ASM is composed of three scanning shadow cameras,
each of which is a position sensitive xenon proportional counter that views
a $6^{\circ} \times 90^{\circ}$\,FWHM section of sky through a
one-dimensional coded mask.  Each camera has a net active area for
detecting X-rays of $\sim$$30\,\rm cm^{2}$. It surveys about 80\% of the
sky 5--10 times each day for $\sim$100\,s. The overall energy range of the
ASM is 1.5--12\,keV which is telemetered in three energy channels
(1.3--3.0\,keV, 3.0--5.0\,keV, and 5.0--12.2\,keV).  Fig.~\ref{fig:asm}b
shows the RXTE ASM counting rate for Cen~A from the beginning of the RXTE
mission, rebinned in 14\,d bins; the arrows denote the times when the
PCA/HEXTE monitoring campaigns were made. We used the ``Data by dwell
quick-look results'' provided by the RXTE ASM team, available from MIT ASM
Light Curve Overview Webpages. We took only flux solutions for which $\chi
^{2}_{\rm red} < 1.2$, with a background estimated between 1 and
10\,counts\,$\rm s^{-1}$ \citep{levine:00a}, and with a full dwell-exposure
time (i.e., 90\,s) .

Also shown in Fig.~\ref{fig:asm}c are the 20--100\,keV count rates from the
BATSE instrument on board the Compton Gamma Ray Observatory (CGRO) averaged
over two week intervals from 1991-1999 (data provided by C.  Wilson,
priv.\ comm.). BATSE consists of eight identical NaI(Tl)
scintillation detector modules located at the corners of CGRO.  Sources are
detected using the Earth occultation technique. During each CGRO orbit, the
Earth is seen by BATSE to sweep across a band in the sky extending
$\sim$$35^{\circ}$ above and below the orbit plane. As each source enters
into (exits from) occultation by the Earth, count rates in the
source-facing detectors decreases (increase) according to the source
intensity.

\section{Data Analysis}\label{sec:dataanalysis}

\subsection{Spectral Analysis}\label{subsec:spectrum}

The PCA and HEXTE data were fit simultaneously using XSPEC v.~11.0q
\citep{arnaud:96a}. The spectrum was modelled by a heavily absorbed power
law plus a Gaussian emission line model, of the form
\begin{eqnarray}\label{eq:model}
  N_{\rm ph}(E) &=& A_{\rm Det} \Bigg[ 
     \mathrm{e}^{-\sigma_\mathrm{ISM}(E) N_\mathrm{ H}} A_\mathrm{ PL}
     E^{-\Gamma} +{}  \nonumber\\   
      && {}+  \frac{\mathrm{A}_\mathrm{Line}}{\sqrt{2 \pi \sigma^2}}
     \mathrm{e}^{-\frac{ \left(E - E_\mathrm{Fe} \right)^2}{2 \sigma^2}}
     \Bigg] 
\end{eqnarray}
To account for the uncertainty in the effective areas of the three detector
systems (the PCA and the two HEXTE clusters), we included their relative
normalizations, $A_{\rm Det}$, as a fit parameter, setting the PCA
normalization to unity. For the photoelectric absorption of the
interstellar medium, $\sigma_{\rm ISM}$, we used the photo-electric
absorption cross section of \citet{morrison:83a} where $ N_{\rm H}$ is the
equivalent hydrogen column. The fit parameters of the photon power law are
$\Gamma$ the photon index and $A_{\rm PL}$ the power law norm ($\rm
ph\,keV^{-1}\,cm^{-2}\,s^{-1}$ at 1 keV).  To model the iron line component
a Gaussian line profile was added where $E_{\rm Fe}$ is the line energy,
$A_{\rm Line}$ the total $\rm ph\,cm^{-2}\,s^{-1}$ in the line, and
$\sigma$ the Gaussian line width. As preliminary fits with the iron line
width as a free parameter resulted in widths narrower than the spectral
resolution of the PCA, we fixed the line width to $\sigma=0.2$\,keV.

The spectral model was fitted to each data set separately (the 1996, 1998,
and 2000 separated spectra) and to the three data sets simultaneously (the
joined spectrum). Table~\ref{tab:spec2} provides fitted parameters and
their associated uncertainties.  Our adopted simple spectral model
(Eq.~\ref{eq:model}) provides a fully acceptable fit in each case. Note
that we assume that the iron line is not absorbed by the intervening
material. This physically implies that the reprocessing region in which the
Fe line is produced is outside the absorbing line of sight, and thus, the
high absorption column is very near the X-ray emission region at the core
of the AGN. This assumption only affects the inferred flux of the line and
its computed equivalent width.

For all three observations, the energy of the iron line, $\sim$6.4\,keV,
indicates that the emission is from cool, i.e., at most weakly ionized
material. It is either possible, 1) that the Fe-line emitting material lies
directly along our line of sight, and could thus be associated with matter
producing the observed $N_{\rm H}$, or 2) it is possible that the Fe-line
emitting material is caused by fluorescent emission of a larger spatial
region. In the former case, i.e., the emission of line photons from within
our line of sight, we would expect the line flux to be correlated with
$N_{\rm H}$. Within the uncertainty of the fit parameters, such a
correlation is not found, as there is no variability in the line flux with
observation while $N_{\rm H}$ varies. Such a constancy of the line flux is
only possible if the line emission originates outside of our line of sight,
for example within the torus.  Here, the line strength would not correlate
with the variation of $N_{\rm H}$. Furthermore, we would also not expect a
correlation with changes of the source flux since these would be smeared
out on the light travel timescale, which can be several years. Note that
the equivalent width of the line decreases between the observations.  This
decrease, however, is purely due to the factor of 1.6 increase in the
continuum flux over the observations. Finally, we note that our
conclusion, that the Fe-line emitting material lies outside of our direct
line of sight, is consistent with estimates for the line strength caused by
the absorbing material alone, which predict that $<20$\% of the line flux
is caused by fluorescence in the absorbing material (see
Sect.~\ref{sec:conclusions}).

Apart from the changes in the power law normalization ($+11\%$ with respect
to 1996) and in the $N_{\rm H}$ ($-30\%$ with respect to 1996) there is no
significant changes of the spectral parameters between the three
observations. In passing, we note that the same holds also true when
analyzing the spectra observed during each individual orbit. No significant
variation of the spectral parameters was found within each monitoring
observation.

\begin{figure}
  \resizebox{\hsize}{!}{\includegraphics{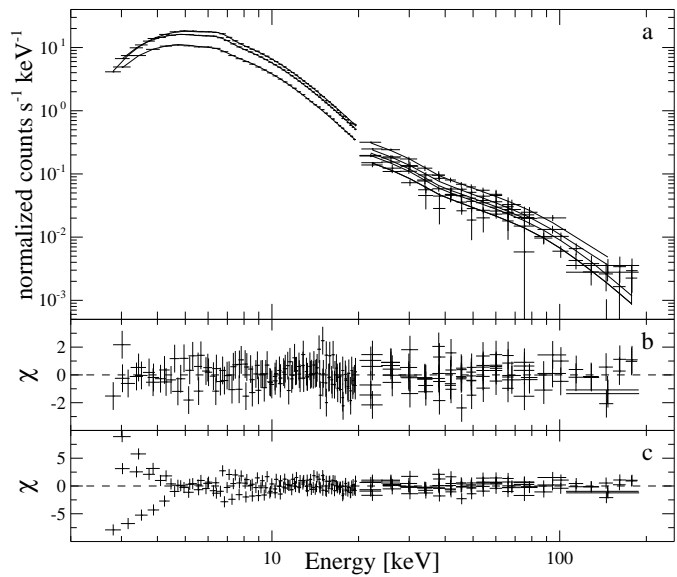}}
  \caption{{\textbf{a}} Best fit joined spectrum for an absorbed power law
    plus Gaussian emission line with $\sigma = 0.2$\,keV.  See
    Table~\ref{tab:spec2}, joined spectrum, for the exact values of the fit
    parameters.  See text for explanations. {\textbf{b}} Residuals to the
    fit in terms of $\sigma$ with error bars of size one.  {\textbf{c}}
    Residuals to the fit with coupled spectral parameter for the three RXTE
    observations.}\label{fig:spectrum}
\end{figure}

As the spectral parameters remain roughly constant between the
observations, it is possible to further constrain the shape of the spectral
continuum by fitting all data simultaneously. We first attempted to fit all
the data with a single model by coupling all spectral parameters (with the
exception of the power law flux) for the three data sets. Our fit to this
model had strong residuals in the region below 10\,keV (see
Fig.~\ref{fig:spectrum}c), clearly indicating the variation in $N_{\rm H}$.
The fit corroborates that the shape of the underlying power-law continuum
might be constant.  We therefore modeled the combined spectrum using a
model where all parameters were allowed to vary freely with the exception
of the photon index, which was held fixed for the three data sets.  The
fitted parameters for this fit are listed in Table~\ref{tab:spec2} (joined
spectrum), a plot of the spectral model and its residuals is shown in
Figs.~\ref{fig:spectrum}a and ~\ref{fig:spectrum}b.  The good $\chi^2$ of
this fit indicates that $\Gamma$ can indeed be assumed to be constant for
the three observations.

In order to test for the presence of a Compton reflection component, a
power law plus Compton component \citep[XSPEC model
{\texttt{pexrav}},][]{magdziarz:95a} fit was performed to both, the
individual observations and the joined data with a single power law index.
The $2\sigma$ upper limit for the reflection component, $\Omega/2\pi$,
expressed as the ratio of the solid angle for primary flux scattered into
our line of sight to an infinite slab, was found to be $0.09$, $0.09$,
$0.08$, and $0.05$ for observations~1, 2, 3, and joined, respectively,
assuming an inclination angle of $70^\circ$.  This result clearly implies
that no significant reflection features are present, confirming the earlier
OSSE and RXTE results \citep{rothschild:99a,kinzer:95a}.

We also tested the RXTE data for the presence of a spectral break at higher
energies, using an exponentially cutoff model (XSPEC model
{\texttt{cutoffpl}}).  The best fit model gave a cutoff-energy with a lower
limit of $500$\,keV and provides an equally adequate fit to the data as the
straight power law ($\chi^{2}_{\rm red}= 0.65$).  This result is consistent
with CGRO OSSE data, where a cutoff energy of $300$\,keV or higher was
found \citep{kinzer:95a}. We would not be sensitive to a break of
$\Delta\Gamma = 0.5$ at 150\,keV, as reported by \citet{steinle:98a}, due
to limited statistics above 150\,keV.

To summarize, the RXTE spectrum of Cen~A can be best described by a heavily
absorbed power law extending above 200\,keV and an iron line from cold
matter. A significant width to the iron line was not required, and
therefore it was taken to be narrow ($\sigma=0.2$\,keV).  No significant
measurements of a steepening in the power law at high energies was
detected. A Compton reflection component was not required to fit the data
with $\Omega/2\pi \leq$ 0.05.

A bright X-ray transient $2\farcm 5$ south-west to the nucleus of Cen~A ,
\object{1RXH J132519.8-430312}, was detected by \citet{steinle:00a} during
ROSAT HRI observations in July 1995.  Parallel multi-wavelength
observations from radio to $\gamma$-ray of Cen~A were made
\citep{steinle:99a} and no counterpart was found in the optical images.
Unfortunately, no spectral information is available from the observations,
and only indirect information can be derived from the BATSE and OSSE
instruments, which provide evidence that the transient emits mainly at soft
X-rays. \citet{steinle:00a} suggested the transient could be an X-ray
binary located in Cen~A. Assuming a power-law spectrum with photon index
$1.5$ and a $N_{\rm H}=8\times 10^{20}\,\rm cm^{-2}$, and taking the ROSAT
HRI average count rate of 0.033\,counts\,$\rm s^{-1}$ obtained by
\citet{steinle:00a}, we estimate a RXTE PCA count rate of
0.36\,counts\,$\rm{s^{-1}\,PCU^{-1}}$ (using the PIMMS tool supported by
HEASARC). This represents $< 2\%$ of the measured Cen~A PCA count rate, and
therefore no significant contribution of the transient source (if it is
supposed to be turned on) on the RXTE observations is expected.  This is
also confirmed by the spectral behavior of the data, where no excess
softness is presented in the spectral fit (see Fig.~\ref{fig:spectrum}).

\subsection{Timing Analysis}\label{subsec:timing}

Variability is one of the well known features of Cen~A and is observed in
all wavelength regimes from radio to $\gamma$-rays
\citep{abraham:82a,turner:92a,terrell:86a,bond:96a,kinzer:95a}. At X-ray
wavelengths, Cen~A is known to be highly variable on time scales of years,
months, and days \citep{jourdain:93a,baity:81a}. This variability can be
split in two main components \citep{jourdain:93a}: (1) long-term trends
which last for years and define the state of the source and (2) short-term
flares of a few days which are superposed on the long-term component.

\subsubsection{Short Term Variability}

To characterize the short-term variability of Cen~A during the PCA/HEXTE
observations we performed a standard $\chi^2$-test for the light curves
on different time scales. A reduced $\chi^2_\nu$ value was computed for
each stream of data to test for variability against the null hypothesis
that the flux remains constant:
\begin{equation}\label{eq:chi2}
\chi^2_\nu = \frac{1}{\nu} \sum_{i=1}^n \frac{(c_i - \langle c
  \rangle)^2}{\sigma_i^2}
\end{equation}
where $c_i$ is the observed count rate at each time interval $i$,
$\sigma_i$ the respective Poisson error, $\langle c \rangle$ the mean count
rate during the observation, $n$ is the number of time bins, and $\nu =
n-1$ is the number of degrees of freedom. For a non variable source we
expected $\chi^2_\nu \sim$$1$.  No variability within the measurement errors
over timescales of days or less was detected for all the pointing
observations of RXTE.  Variations of $10\%$ in the mean flux were detected
over timescales of $\sim$$5$\,d during observation~2 (see
Fig.~\ref{fig:pca_lc}, middle panel).

\begin{figure}
  \resizebox{\hsize}{!}{\includegraphics{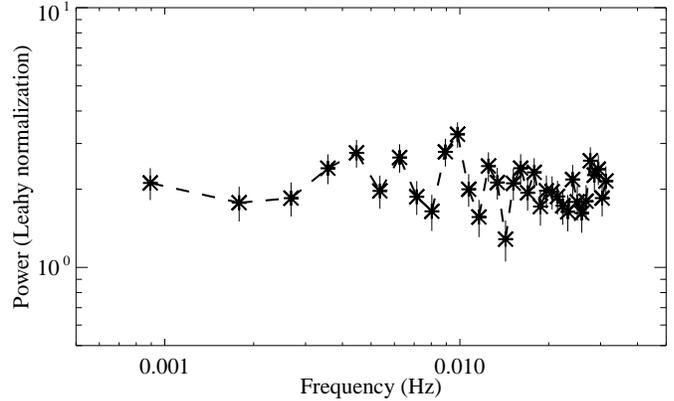}}
  \caption{ Averaged power spectral density (PSD) in \citet{leahy:83a}
    normalization of the PCA background subtracted source data. A PSD value
    of 2 corresponds to pure Poisson noise.}\label{fig:psd}
\end{figure}

Fourier techniques were also used for the analysis of the long 1998
observation. An average power spectral density (PSD) using time segments of
$\sim$$22$\,min duration was calculated. This duration was the longest time
span for which enough time segments without gaps were available to compute
a PSD. The analysis was performed for the PCA background subtracted data
from 4 to 40\,keV, where the data is source dominated. No excess
variability above the Poisson level was seen (see Fig.~\ref{fig:psd}),
confirming the result of the $\chi^2$ test in this frequency. 

To compare the Cen~A variability with previous studies of Seyfert Galaxies
\citep{markowitz:00a,turner:97b,nandra:97a} we extracted the 2--10\,keV
light curve of the 1998 monitoring campaign and computed the normalized
``excess variance'', $\sigma^2_{\rm rms}$ \citep[for a detailed discussion
of the definition see][]{turner:99a,nandra:97a}:
\begin{equation}\label{eq:rms}
\sigma^2_{\rm rms} = \frac{1}{n \langle c \rangle^2} \sum_{i=1}^n [(c_i -
\langle c \rangle )^2 - \sigma^2_i]
\end{equation}
We find $\sigma^2_{\rm rms}=4.3(9) \times 10^{-3}$. Previous studies have
found a clear decrease of $\sigma^2_{\rm rms}$ with the intrinsic X-ray
luminosity of both Seyfert~1 and Seyfert~2 galaxies \citep[consistent also
with the time-domain results of][and with \citealt{green:93a}]{koenig:97b}.
Our result for Cen~A is a factor of $\sim$10 smaller than that expected
from the Seyfert correlations \citep{turner:97b,nandra:97a} for the
intrinsic (i.e., unabsorbed) X-ray luminosity of Cen~A ($\log L_{\rm X,
  2-10\,\rm keV} \simeq 41.7\,\rm cgs$). Such a low $\sigma^2_{\rm rms}$ is
typically only expected for Seyfert~2's with $\log L_{\rm X}\sim$$43\,\rm
cgs$ \citep[see][Fig.~2]{turner:97b}, although the scatter in these data is
large. We conclude that the short term variability of Cen~A is on the low
side of that observed in comparable Seyfert galaxies.

\begin{figure*}
  \resizebox{\hsize}{!}{\includegraphics{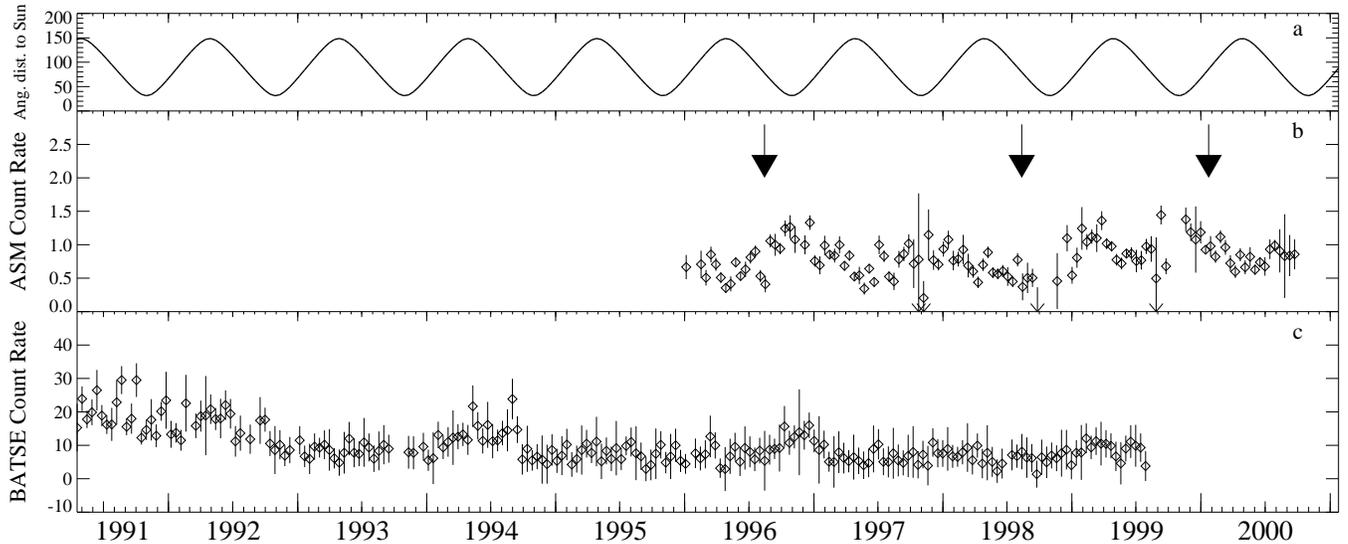}}
  \caption{{\textbf{a}} Angular distance between the Sun and
    Cen~A. {\textbf{b}} ASM counting rate for Cen~A from the beginning of
    the RXTE measurements, rebinned in 14\,d bins. The RXTE PCA/HEXTE
    monitoring campaigns are indicated by the arrow.  {\textbf{c}} BATSE
    counting rate for Cen~A from the beginning of the CGRO
    measurements, rebinned in 14\,d bins.}\label{fig:asm}
\end{figure*}

\subsubsection{Long Term Variability}

To study the long-term variability of Cen~A, and to place our pointed
observations within the context of the overall behavior of the source, we
used data from the RXTE ASM and CGRO BATSE all sky monitors (see
Fig.~\ref{fig:asm}).

Fig.~\ref{fig:asm}b indicates apparent fluctuations in the ASM flux at the
end of each year coinciding with the closest angular separation between the
sun to Cen~A (Fig.~\ref{fig:asm}a).  This variability is due to scattering of
solar X-rays off the ASM collimators onto the ASM detectors and is not due
to source flux variations (Remillard, priv.\ comm.).

The BATSE count rate is shown in Fig.~\ref{fig:asm}c and does not display
the variability seen in the ASM.  Thus, we conclude that the Cen~A flux
appears to have varied by less than a factor of two during the four years
of the RXTE mission. BATSE data extending back to 1991 indicate that the
same conclusion can be drawn back to at least mid-1992.  Cen~A was a factor
of two brighter in the BATSE data from mid-1991 to mid-1992.

\section{Discussion and Conclusions}\label{sec:conclusions}

We have presented analysis of three RXTE monitoring campaigns of Cen~A,
totaling $\sim$$110$\,ks. This analysis reveals no significant short-term
temporal variability on timescales from weeks or less, and a relatively
simple spectrum in the 2--200\,keV range. The best fit spectral model
consists of a heavily absorbed primary flux and a narrow emission line from
cool iron. The iron line flux remained stable at $\sim$$\rm 4.5 \times
10^{-4}\,photons\,cm^{-2}\,s^{-1}\,keV^{-1}$ for unabsorbed flux and
$\sim$$\rm 5.3 \times 10^{-4}\,photons\,cm^{-2}\,s^{-1}\,keV^{-1}$ for
absorbed flux.

The 2--10 keV flux was the same in 1996 and 1998, and increased by 60\% in
2000. This implies that the line emitting region is not very close (greater
than light days?) to the region containing the primary flux generation.
This is consistent with the lack of a significant reflection component
contribution to the overall flux.  We can conclude that the Fe line does
not originate close to the central X-ray source, although a strong Fe line
is observable.  Possible sources of the Fe line flux would then be a
molecular torus surrounding the source (similar to Seyfert~2 galaxies) or
the dust lane.  To test the latter hypothesis we have used a Monte Carlo
code to compute the iron line flux emerging from the back of a slab of
neutral material with $N_{\rm H}=7\times 10^{22}\,\rm cm^{-2}$ that is
irradiated by the unattenuated continuum spectrum of Cen~A.  The equivalent
width of the line resulting from the absorbing material is $<30$\,eV so
that most of the contributing Fe line photons do not come from the
absorbing region.  In agreement with \citet{grandi:99a} and
\citet{wozniak:98a} this leaves us with the conclusion that the Fe line
originates within material of a large column outside of the line of sight,
for example in the postulated molecular torus.  Our conclusion is also
confirmed by the presence of no correlation between the variability
behavior of $N_{\rm H}$ and the iron line flux, since $N_{\rm H}$ varied
with observation and the line flux does not. The $N_{\rm H}$ could
originate at the edge of the accretion disk and thus its variability might
be due to small changes in the disk structure, while the iron line material
lies outside the line of sight.

The lack of a reflection component in radio-loud active galaxies poses a
problem for unification models of AGN. There are several ways in which a
reflection-free AGN spectrum could be produced. First, the reflecting cold
material could be removed from the inner regions by being ``swept up'' by
the radio jet. For Cen~A, there are indications that such a mechanism might
be at work. In a recent interpretation of Hubble Space Telescope images
taken with WFPC2 and NICMOS, \citet{marconi:00a} notice an evacuated
channel produced by the jet between the nucleus and the first bright knot
of the jet. A second possibility might be that the inner regions of the
accretion disks of radio-loud AGN are not neutral, but strongly ionized.
The source for the ionization could be radiation of the radio-jet, or the
accretion disk structure could deviate from that of a simple accretion
disk, and be, e.g., advection dominated (the latter is improbable, however,
because of to the appreciable X-ray luminosity of the source). Lastly, it
could also be possible that the production site of the primary X-rays is
not within an accretion disk but within the base of the jet itself, such
that the covering factor of any cold material present close to the central
black hole is small. Observationally, this last interpretation has been
advocated by \citet{kinzer:95a} in the analysis of CGRO OSSE data from
Cen~A, where the energy dependence of the high energy cutoff of the
power-law with source flux is interpreted in terms of a changing maximum
energy of a nonthermal electron distribution that would be expected in the
jet.

\begin{acknowledgements}
  We acknowledge the RXTE Science Operations Center staff providing the
  observations and the Guest Observer Facility for providing support in
  analyzing them. The research has been partially financed by {\emph{la
      Caixa}}/DAAD grant A/98/19182, by NASA grant NAS5-30720, by
  international NSF travel grant INT-9815741, and by a travel grant from
  the DAAD. SB and JW acknowledge the hospitality and financial support of
  the Center for Astrophysics and Space Sciences of the University of
  California at San Diego. RR and WH similarly acknowledge the friendly
  atmosphere of the Institut f\"ur Astronomie und Astrophysik of the
  University of T\"ubingen. CSR appreciates support from Hubble Fellowship
  grant HF-01113.01-98A. This grant was awarded by the Space Telescope
  Institute, which is operated by the Association of Universities for
  Research in Astronomy, Inc., for NASA under contract NAS 5-26555.
\end{acknowledgements}


\end{document}